\documentclass[11pt,preprint]{aastex}
\usepackage{natbib}
\usepackage{amsmath}
\usepackage{xspace}
\usepackage{xcolor}
\newcommand{\beq}[1]{\begin{equation}\label{#1}}
\newcommand{\eeq}{\end{equation}}

\newcommand{\highlight}[1]{\textcolor{black}{#1}\xspace}

\shorttitle{Fast Radial Flows}
\shortauthors{Rosenfeld, Chiang, \& Andrews}

\begin{document}

\title{FAST RADIAL FLOWS IN TRANSITION DISK HOLES}

\author{Katherine A. Rosenfeld\altaffilmark{1}, Eugene Chiang\altaffilmark{2},
        and Sean M. Andrews\altaffilmark{1}}
\altaffiltext{1}{Harvard-Smithsonian Center for Astrophysics, 60 Garden Street, Cambridge, MA 02138, USA}
\altaffiltext{2}{Departments of Astronomy and Earth and Planetary Science, University of California, Berkeley, CA 94720, USA}

\begin{abstract}

Protoplanetary ``transition'' disks have large, mass-depleted central cavities,
yet also deliver gas onto their host stars at rates comparable to disks
without holes.  The paradox of simultaneous transparency and accretion can be 
explained if gas flows inward at much higher radial speeds inside the cavity 
than outside the cavity, since surface density (and by extension optical depth) 
varies inversely with inflow velocity at fixed accretion rate.  Radial speeds 
within the cavity might even have to approach free-fall values to explain the 
huge surface density contrasts inferred for transition disks.  We identify 
observational diagnostics of fast radial inflow in channel maps made in 
optically thick spectral lines.  Signatures include (1) twisted isophotes in 
maps made at low systemic velocities and (2) rotation of structures observed 
between maps made in high-velocity line wings.  As a test case, we apply our 
new diagnostic tools to archival ALMA data on the transition disk HD 142527, 
and uncover evidence for free-fall radial velocities inside its cavity. 
%EC
\highlight{Although the observed kinematics are also consistent with a
  disk warp, the radial inflow scenario is preferred because it
  predicts low surface densities that appear consistent with recent
  observations of optically thin CO isotopologues in this disk.}
%Nevertheless the observed kinematics are also consistent with a disk warp;
%disentangling the two scenarios calls for detecting optically thin line 
%emission to measure the low surface densities expected to accompany fast 
%inflow.  
How material in the disk cavity sheds its angular momentum wholesale 
to fall freely onto the star is an unsolved problem; gravitational torques 
exerted by giant planets or brown dwarfs are briefly discussed as a candidate 
mechanism.
\end{abstract}


\section{INTRODUCTION} \label{intro}

A subset of T Tauri and Herbig Ae disks have large, mass-depleted cavities in 
their inner regions, as inferred from their infrared spectra 
\citep[e.g.,][]{Kim13} or spatially resolved imaging at infrared 
\citep[e.g.,][]{Geers2007} and mm wavelengths \citep[e.g.,][]{Brown09}.  The
cavities in these ``transition'' disks have radii ranging from $\sim$4\,AU 
\citep[TW Hya;][]{Calvet02,Hughes07} to 20\,AU \citep[GM Aur;][]{hughes09} to 
40--50\,AU \citep[LkCa 15;][]{Andrews11,Isella12} to 130--140\,AU \citep[HD 
142527;][]{casassus13}.  The suspicion is that dynamical interactions with 
nascent planetary systems evacuate the inner disk regions 
\citep[e.g.,][]{Zhu11}.

Surprisingly, despite their optically thin cavities, many transition
disks appear to be accreting gas at fairly normal rates.  Stellar
accretion rates $\dot{M}_\ast$ in transition disks range from
$\sim$$10^{-9}\,M_\odot\,{\rm yr}^{-1}$ \citep[TW Hya;][]{muzerolle00}
to $10^{-8} \,M_\odot\,{\rm yr}^{-1}$ \citep[GM
Aur;][]{gullbring98,Calvet05} to $10^{-7}\, M_\odot\,{\rm yr}^{-1}$
\citep[HD 142527;][]{garcia06}.  These rates overlap with those for
conventional disks without cavities, although the median
$\dot{M}_\ast$ for transitional disks may be a factor of $\sim$10
lower than for conventional disks
\citep{najita07,espaillat12,Kim13}.\footnote{\highlight{Of course, there are
  disks, transitional and otherwise, with lower accretion rates than
  those cited in the main text (e.g., \citealt{sicilia-aguilar10};
  \citealt{rigliaco12})---at least insofar as attempts to measure
  accretion rates can be trusted (e.g., \citealt{demarchi10};
  \citealt{curran11}; \citealt{alcala13}). Some transitional disks
  with unmeasurably low stellar accretion rates may harbor stellar binaries
  whose torques are strong enough to stave off accretion (e.g., CoKu
  Tau/4; \citealt{ireland08}; \citealt{najita07}).  Our focus in this
  paper is on healthily accreting transitional disks, as they pose the
  greatest paradox: how can they have cavities and accrete at the same
  time?}}

It is sometimes thought that transition disk cavities are created by planets 
consuming material from the outer disk and thereby starving the inner disk.  
This is unlikely to be correct --- or at least it cannot be the whole story --- 
because a starved inner disk would also starve the host star, contrary to the 
observed values of $\dot{M}_\ast$ cited above.  Modeling of disk spectra and 
images reveals that surface density contrasts inside and outside cavities can 
be $\gtrsim 10^3$ \citep[e.g.,][]{Calvet05}.  Planets inside the cavity cannot
simultaneously eat $\gtrsim 99.9$\% of the mass flowing inward from the outer 
disk and still leave enough for their host star to accrete at near-normal 
rates.  If we interpret the factor-of-10 lower $\dot{M}_\ast$ for transition 
disks as arising from planets consuming $\sim$90\% of the accretion flow 
from the outer disk and leaving the remaining $\sim$10\% for the central star 
\citep[e.g.,][]{lubow06}, then the surface density just inside the planet's 
orbit would decrease by only a factor of 10, which is almost certainly too 
small to render the inner disk optically thin --- all other factors being 
equal.

In this paper, we describe a way for transition disks to have their cake 
(possess optically thin cavities) and eat it too (accrete at normal rates).  
The dilemma can be resolved by (somehow) increasing the radial accretion 
velocity $v_R$ inside the cavity, thereby concomitantly lowering the surface 
densities $\Sigma$ there.  For a given radius $R$ and steady disk accretion 
rate $\dot{M} = 2\pi R \Sigma v_R$, we have $\Sigma \propto 1/v_R$.  Just 
outside the cavity, $\Sigma$ is large because accretion in the outer disk is 
diffusive (``viscous") and slow ($v_R \sim \alpha (H/R)^2 v_{\rm K} \ll v_{\rm
 K}$, where $H$ is the disk vertical thickness, $\alpha \ll 1$ is the 
Shakura-Sunyaev stress parameter, and $v_{\rm K}$ is the Keplerian orbital 
velocity).  Just inside the cavity, the accretion velocity $v_R$ is somehow 
boosted --- $v_R$ could, in principle, approach its maximum of 
$\sim$$v_{\rm K}$ --- so that $\Sigma$ is lowered, conceivably by orders of 
magnitude, while maintaining the same mass flow rate $\dot{M}$ as in the outer 
disk.  

Gravitational (as opposed to magnetic or hydrodynamic) torques are an obvious 
candidate for catastrophically removing angular momentum from gas inside the 
cavity.  In galaxies, gravitational torques are exerted by stars that arrange 
themselves into non-axisymmetric patterns.  A stellar bar (a perturbation with
azimuthal wavenumber $m=2$) eviscerates the interior of the $\sim$3-kpc 
molecular ring at the center of our Galaxy \citep{Schwarz81,Binney91}; an 
eccentric stellar disk ($m=1$) may be funneling gas onto the star-forming 
nucleus of M31 \citep{Chang07}; and a complex nest of $m=1$ and $m=2$ stellar 
potentials force-feeds gas onto quasars \citep{Shlosman89,Hopkins11}.  In these 
galaxy-scale examples, radial inflow velocities can approach free-fall rates.

The role of stars in galaxies might be played by planets in transition disks.  
We like this interpretation, but it is not without difficulty.  The cavities 
are so large (they are not annular gaps) and so clean that appeals are made to 
rather extreme parameters.  As many as four planets, each weighing up to 
$\sim$10 $M_{\rm Jup}$, have been posited to carve out a given cavity 
\citep{Zhu11,Dodson11}.\footnote{Even multiple super-Jovian planets are 
reportedly not enough: grain growth and dust filtration at the cavity rim are 
invoked in addition to planets to make the cavity sufficiently transparent 
\citep{Zhu11,Dong12}.}  At least in the case of the transition disk around LkCa 
15, there might be one (but so far only one) actively accreting super-Jovian 
planet located well inside the disk cavity \citep{Kraus12}.  Some assistance in 
driving the accretion flow inside the cavity may be obtained from the 
gravitational potential of the outer disk itself, if that disk is sufficiently 
clumpy \highlight{and massive (for examples of clumpy disks, see the 
observations of IRS 48 by \citealt{vandermarel13}, and the studies of HD 142527 
by \citealt{casassus13} and \citealt{fukagawa13}).}  Just inside a 
non-axisymmetric outer disk, gas streamlines can cross, shock, and plunge 
inward \citep[e.g.,][]{pac77}.

Disk radial velocities driven by embedded massive planets or brown dwarfs can 
be healthy fractions of orbital (azimuthal) velocities.  We derive a crude 
estimate\footnote{Our ``zero-dimensional'' analysis ignores the fact that 
actual disk flows around planets are two-dimensional and dominated by
non-axisymmetric streams flowing in and around planetary Hill spheres.
Still, some indirect empirical support for the scalings in equation 
(\ref{eqn:simple}) has been found by \cite{Fung13}.
}
 for $v_R$ by equating the one-sided Lindblad torque --- the amount of 
angular momentum transmitted outward across a planet's orbit 
\citep{goldreich80} --- with the angular momentum advected inward by accretion, 
$\dot{M} v_{\rm K} R$.  This balance yields
\begin{equation}
v_R \sim \left( \frac{M_p}{M_\ast} \right)^2 \left( \frac{R}{H} \right)^3 v_{\rm K} \label{eqn:simple}
\end{equation}
where we have used the total linear Lindblad torque, integrated over all 
resonances outside the torque cut-off zone at distances $\gtrsim H$ away from 
the planet (see, e.g., \citealt{goodman01} or \citealt{crida06}, and references 
therein).  For a planet mass $M_p \sim 10\,M_{\rm Jup}$, a host stellar mass 
$M_\ast \sim 1\,M_\odot$, and a typical disk aspect ratio $H/R \sim 0.1$, 
equation (\ref{eqn:simple}) evaluates to
\begin{equation}
v_R \sim 0.1 \,v_{\rm K} \sim 0.5 \left( \frac{30\,{\rm AU}}{R} \right)^{1/2}\,{\rm km} \,\, {\rm s}^{-1}. 
\label{eqn:simple1}
\end{equation}
Modern heterodyne receivers on (sub)millimeter wavelength interferometers 
routinely observe disks at $\sim$0.1 km s$^{-1}$ resolution, and the Atacama 
Large Millimeter Array (ALMA) is expected to provide high sensitivity data at 
finer velocity scales yet.  Evidence for free-fall radial velocities can be 
found in the HCO$^+$ filament within the disk cavity around HD 142527.  
\citet{casassus13} discovered this filament to be oriented nearly radially; in 
estimating $\dot{M}$ for this system, they assumed $v_R \sim v_{\rm K}$ 
(although the velocities of the HCO$^+$ filament were not actually quoted in
their study).

Regardless of what is actually accelerating the accretion, we ask here whether 
we can test observationally the idea that transition disk cavities have fast 
(nearly free-fall) radial flows.  We present in \S \ref{sec:twist} some 
predicted observational signatures of radial inflow in spectrally and spatially 
resolved (sub)millimeter-wavelength maps of molecular emission lines.  A sample
case study based on the HD 142527 disk is presented in \S \ref{sec:prospects}.  
We conclude in \S \ref{sec:conclude}.

\section{OBSERVATIONAL SIGNATURES OF RADIAL INFLOW} \label{sec:twist}

Interferometers observing the (sub)millimeter and radio spectral line 
transitions of molecular gas tracers produce a spatially resolved map of the 
source for each frequency (velocity) channel.  A disk in Keplerian rotation has 
well-ordered gas velocities, and so line emission in a particular channel
``highlights'' disk material that is appropriately Doppler shifted 
\citep[e.g.,][]{omodaka92,beckwith93}.  Adding a fast radial inflow to the 
normal Keplerian velocity field alters the morphology of channel map emission
\citep[cf.][]{tang12}.  We identify two related observational signatures of 
radial inflow in a disk cavity: (1) twisted isophotes in the channel map
corresponding to the systemic velocity (\S \ref{subsec:isotwist}), and (2) 
rotation of features seen between maps made in high-velocity wings (\S\ref{subsec:wingrot}).
While distinctive, neither of these features is entirely unique, and must be 
disentangled from other non-Keplerian effects (\S \ref{subsec:disentangle}).

Our analysis uses two axisymmetric models: a razor-thin disk and a 
three-dimensional (3D) disk.  The razor-thin model is a pedagogical tool to 
develop intuition.  The 3D model is more realistic and presents a platform to 
test the razor-thin model as well as analyze mock data products.

\begin{figure}[t!]
\epsscale{1.}
\plottwo{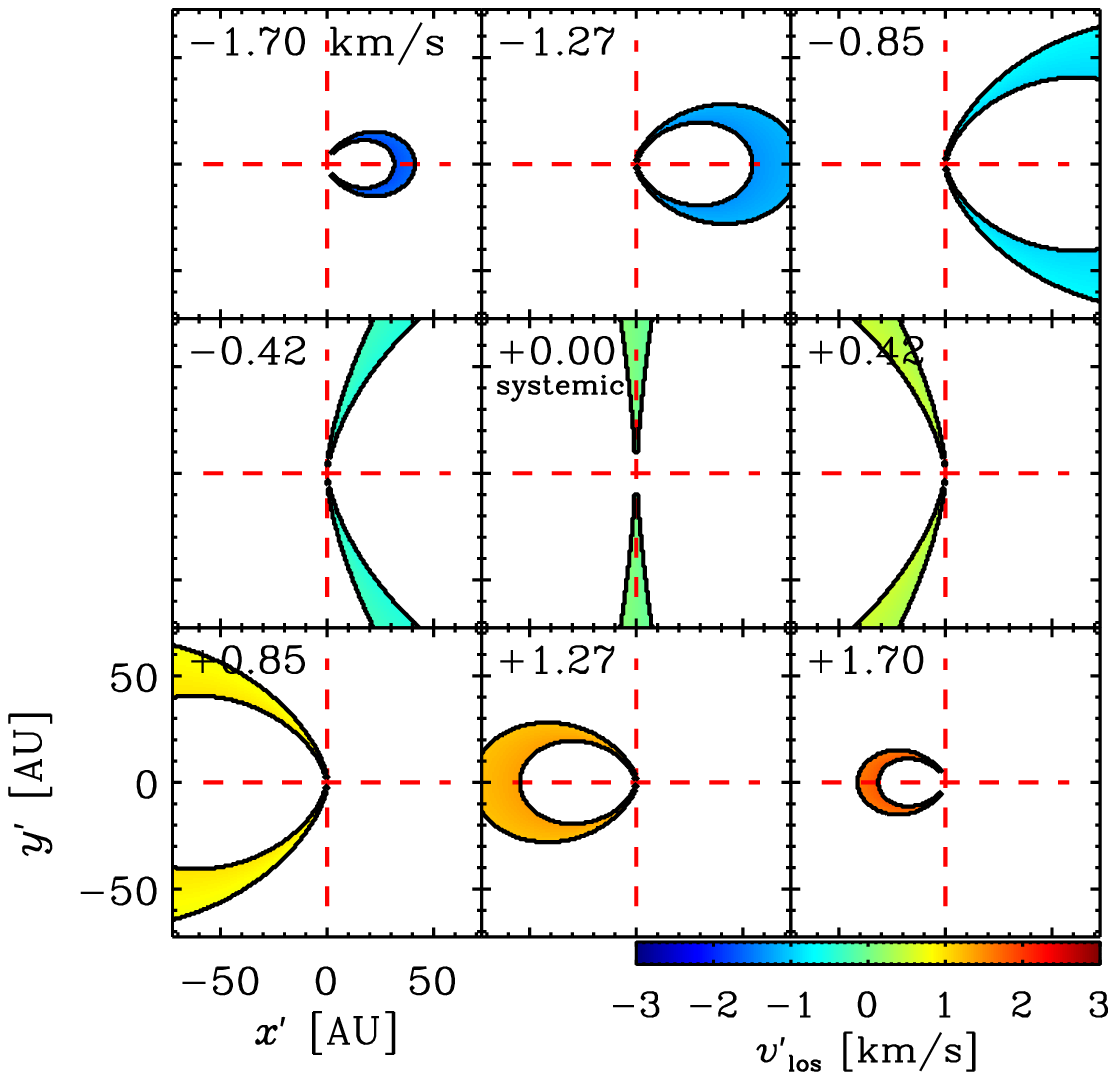}{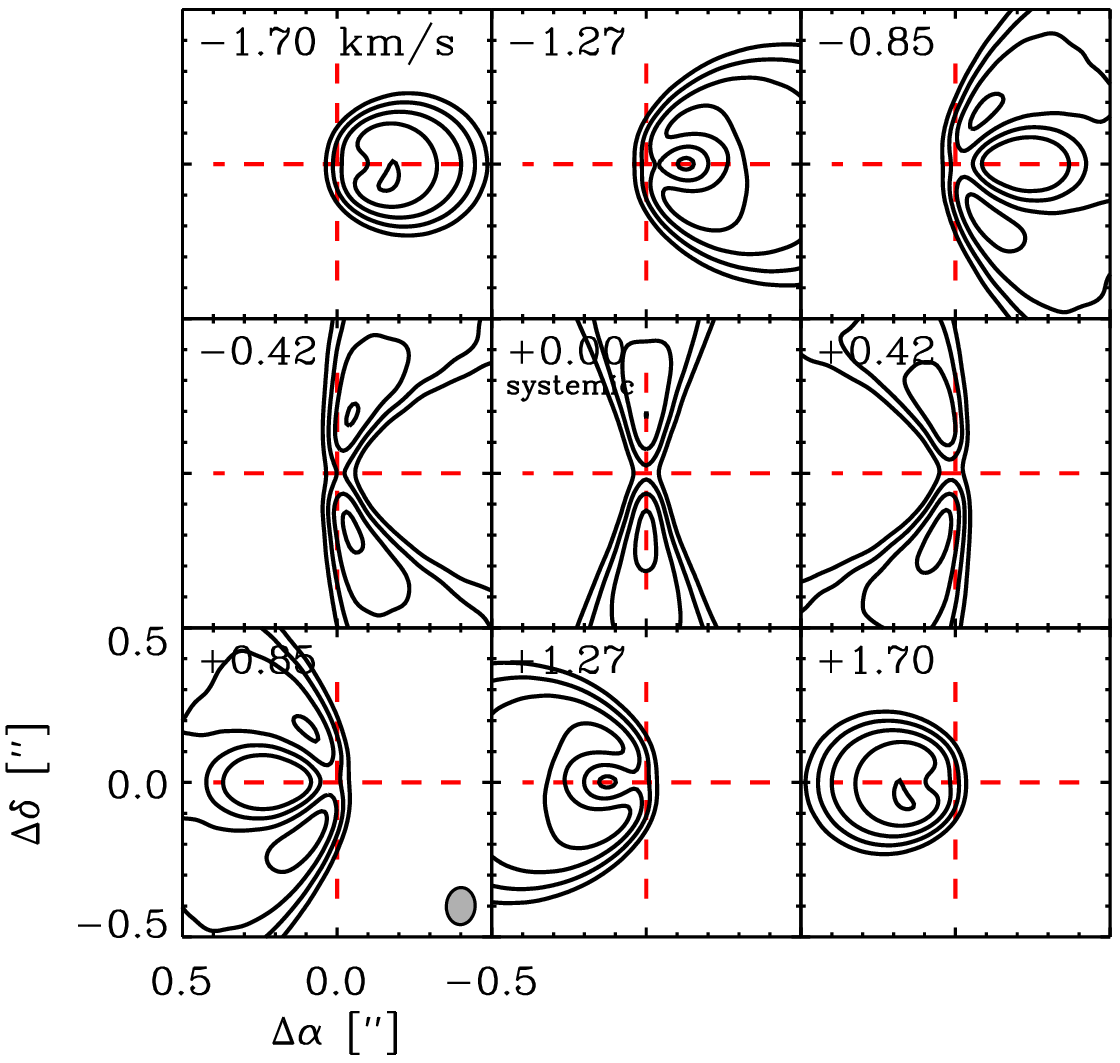}
\figcaption{
{\it Left}: Regions of a razor-thin disk model ($i=20^\circ$, $\chi=0$) where 
the velocities $v'_{\rm los}$ fall within each labeled spectral channel of 
width $\Delta v'_{\rm los} = \pm \,0.212$ km s$^{-1}$.  {\it Right}: Simulated 
$^{12}$CO $J$=3$-$2 channel maps for a 3D disk with the same $i$ and $\chi$.  
Projected major (horizontal) and minor (vertical) axes of the disk are shown as 
dashed red lines.  The spatial scales of all panels are identical for a 
distance of 145\,pc.  The morphology of line emission in the channel maps 
(right) respects the velocity field (left).  The synthesized beam for the 
simulated 3D model is 0\farcs12$\times$0\farcs09, with the 
beam major axis oriented at position angle PA = $-3^\circ$. \label{fig:models}}
\end{figure}

Both models treat the gas velocities within the cavity ($R<R_{\rm cav}$) as 
having an inward radial component that is some constant fraction $\chi$ of the 
local circular Keplerian velocity.  The velocity field is then
\begin{align}
v_x =& -v_{\rm K} \sin\theta - \chi v_{\rm K} \cos \theta \\
v_y =& +v_{\rm K}  \cos\theta - \chi v_{\rm K} \sin \theta \,,
\end{align}
both for the razor-thin model and the 3D model.  Here $\theta$ is the azimuthal 
angle in the disk plane, and variations in bulk velocity with height $z$ are 
ignored (but see \S \ref{subsec:disentangle}).  Outside the cavity, gas is 
assumed to follow circular Keplerian orbits.  Our assumption that the radial 
inflow is axisymmetric is probably unrealistic --- certainly so if the inflow 
is driven by large-scale gravitational torques as we have described in 
\S\ref{intro}.  Nevertheless we assume axisymmetry to gain a foothold on the 
problem.

The disk is inclined by a viewing angle $i$ about the $x$ axis, where 
$i=0\degr$ corresponds to a disk seen face-on.  Note that the effects we will 
discuss are sensitive to the spin orientation of the disk, and so $i$ can range 
between $0\degr$ and $180\degr$ without degeneracy.  Inclinations $0\degr < i < 
90\degr$ imply that the line-of-sight component of the orbital angular momentum 
points toward the observer, whereas $90\degr < i < 180\degr$ implies 
that the line-of-sight component points away.

For the razor-thin model, the projected line-of-sight velocity is\footnote{We 
can set $v'_{\rm los}$  equal to either $+ v_y \sin i$ or $- v_y \sin i$.  The 
degeneracy arises because we can choose the disk's ascending ($\equiv$ toward 
the observer) node on the sky plane to coincide with either the $x=x'$ axis 
(our convention in this paper) or the $-x = -x'$ axis.  Another way to say this 
is that $i$ is degenerate with $-i$.}
\begin{equation}
v'_{\rm los}= -v_y \sin i = -v_{\rm K}\sin i\cos\theta+\chi v_{\rm K} \sin i\sin\theta.
\label{eq:vlos}
\end{equation}
We utilize primed coordinates to denote values measured by the observer; the 
spatial coordinates in the sky plane are $x'=x$ and $y'=y\cos i$.  The azimuth 
$\theta$ measured in the disk plane is related to the on-sky angle $\theta'$ by 
$\tan\theta'= \cos i \tan \theta$.  The left panel of Figure \ref{fig:models} 
shows the morphology of line emission for a razor-thin disk with no inflow.

We compute an analogous 3D disk model that simulates the $^{12}$CO
$J$=3$-$2 line emission from a disk with a total gas mass of
0.01\,$M_\odot$, a spatially constant CO:H$_2$ number-abundance
fraction of $10^{-4}$, and a Gaussian vertical density distribution
($\exp(-z^2/2H^2)$ with a scale height of $H=0.1R$).  The surface
density profile is that of a thin, viscously accreting disk
\citep{bellpringle74}: $\Sigma\propto R^{-1}\exp(-R/R_s)$ with scale
radius $R_s=50$ AU and an outer truncation radius of 300 AU.  Since
our focus in this paper is on the velocity field and not on the
density structure, we have not included any reduction of $\Sigma$
within the cavity.  \highlight{ This simplification should not
  introduce serious error as long as the cavity region remains
  optically thick in $^{12}$CO emission---we estimate that this condition holds
  for $\dot{M} \gtrsim 10^{-8} M_\odot/$yr and $R_{\rm cav}
  \lesssim 50$ AU, even if $\chi$ jumps discontinuously from 0 to 1
  across the cavity boundary.  These requirements are satisfied for HD
  142527, our subject test case of \S\ref{sec:prospects}.}

\highlight{The model is vertically isothermal with a power-law radial 
temperature profile, $T(R) = 25\,(R/100$\,AU$)^{-0.5}$\,K: this temperature 
profile is used only to calculate the level populations of molecules, not the 
hydrostatic disk structure}.  The mass of the central star is 1\,$M_\odot$, 
\highlight{the distance to the observer is 145\,pc}, and the cavity radius is 
$R_{\rm cav} = 50$\,AU.  The disk inclination $i$ and radial inflow parameter 
$\chi$ are variable.  We use the modeling capabilities of {\tt LIME} 
\citep{brinch10} to simulate high-resolution channel maps assuming LTE 
conditions and a turbulent linewidth of $\xi = 0.01$\,km\,s$^{-1}$.  We then 
use {\tt CASA} to sample and deconvolve the Hanning-smoothed visibilities with 
a spectral resolution of 0.212\,km\,s$^{-1}$ and a spatial resolution of 
0\farcs1: \highlight{these parameters should typify
spectral line observations of disks by ALMA when the instrument reaches 
technical maturity}.  The right panel of Figure \ref{fig:models} shows the 
channel maps from a 3D disk model ($\chi=0$), zoomed into the inner 
$\pm$0\farcs5.  Comparing the left and right panels of Figure \ref{fig:models} 
reveals how the velocity field of the razor-thin model offers a guide to the 
morphology of line emission in the channel maps of the 3D model \citep[see 
also][]{beckwith93}.

\begin{figure}[t!]
\epsscale{1.0}
\plotone{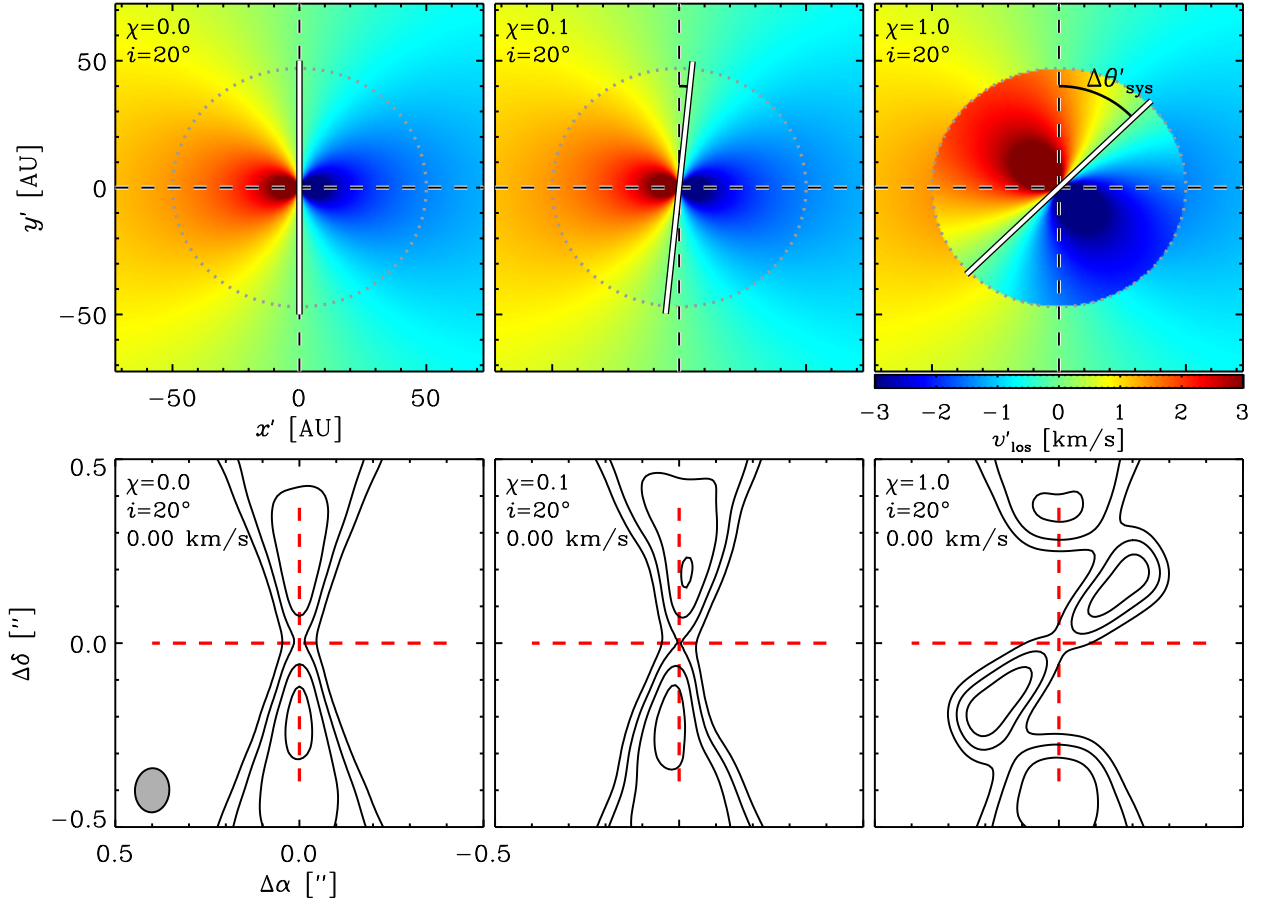}
\figcaption{Twisted isophotes at the systemic velocity. {\it Top}: Observed 
velocity fields $v'_{\rm los}$ for razor-thin disks with varying degrees of 
radial inflow.  Within the cavity radius $R_{\rm cav}=50$\,AU, gas flows 
axisymmetrically inward with a velocity $\chi$ times the local circular 
Keplerian velocity.  The dotted gray circle traces the cavity boundary, the 
solid white bar indicates $\theta'_{\rm sys}$, and the black arc traces $\Delta 
\theta'_{\rm sys} = \theta'_{\rm sys} - 90^\circ$.  Dashed lines show the 
projected major (horizontal) and minor (vertical) axes.  {\it Bottom}: 
Simulated channel maps at the systemic velocity for 3D disks having the same 
$i$ and $\chi$ as the razor-thin models shown in the top panels.  The isophotes 
twist at the location of the cavity radius ($\sim$0\farcs34 from the map 
centers). \label{fig:thetasys}}
\end{figure}

\begin{figure}[t!]
\epsscale{1.0}
\plottwo{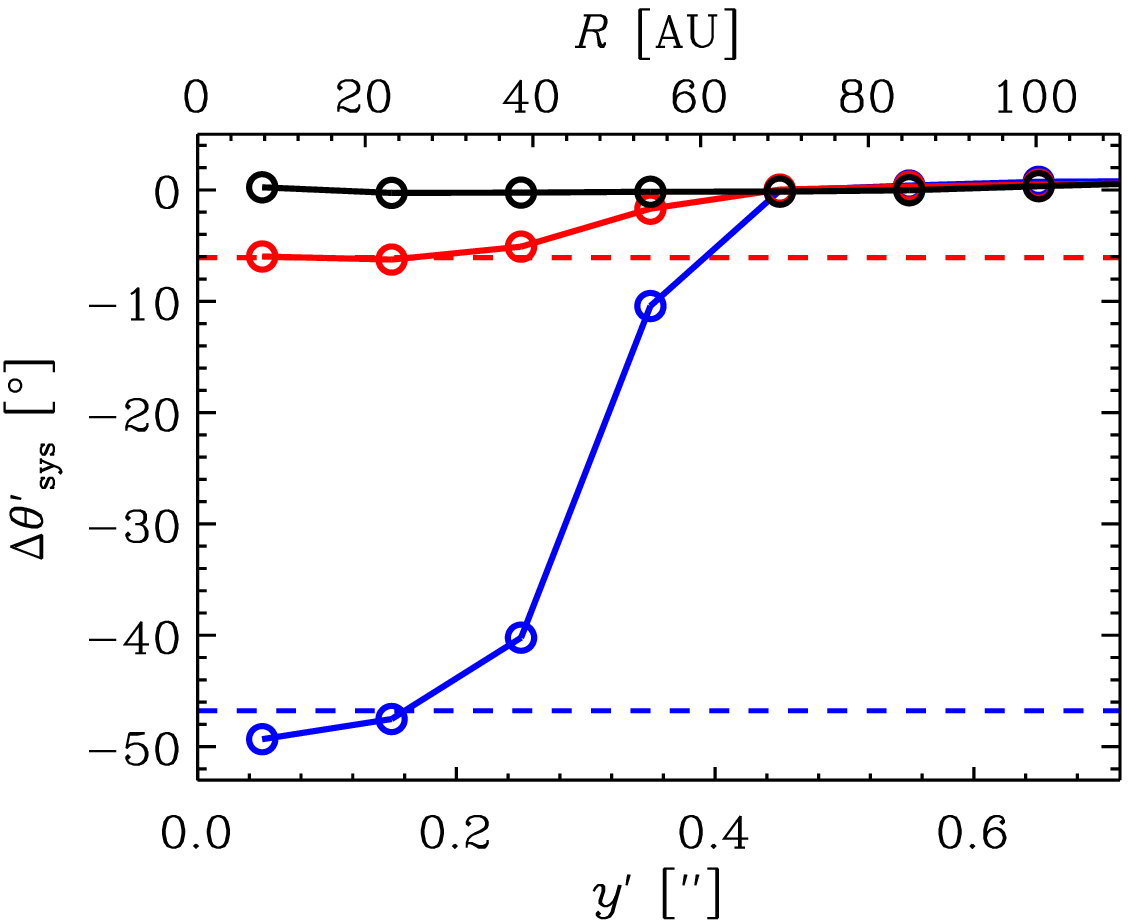}{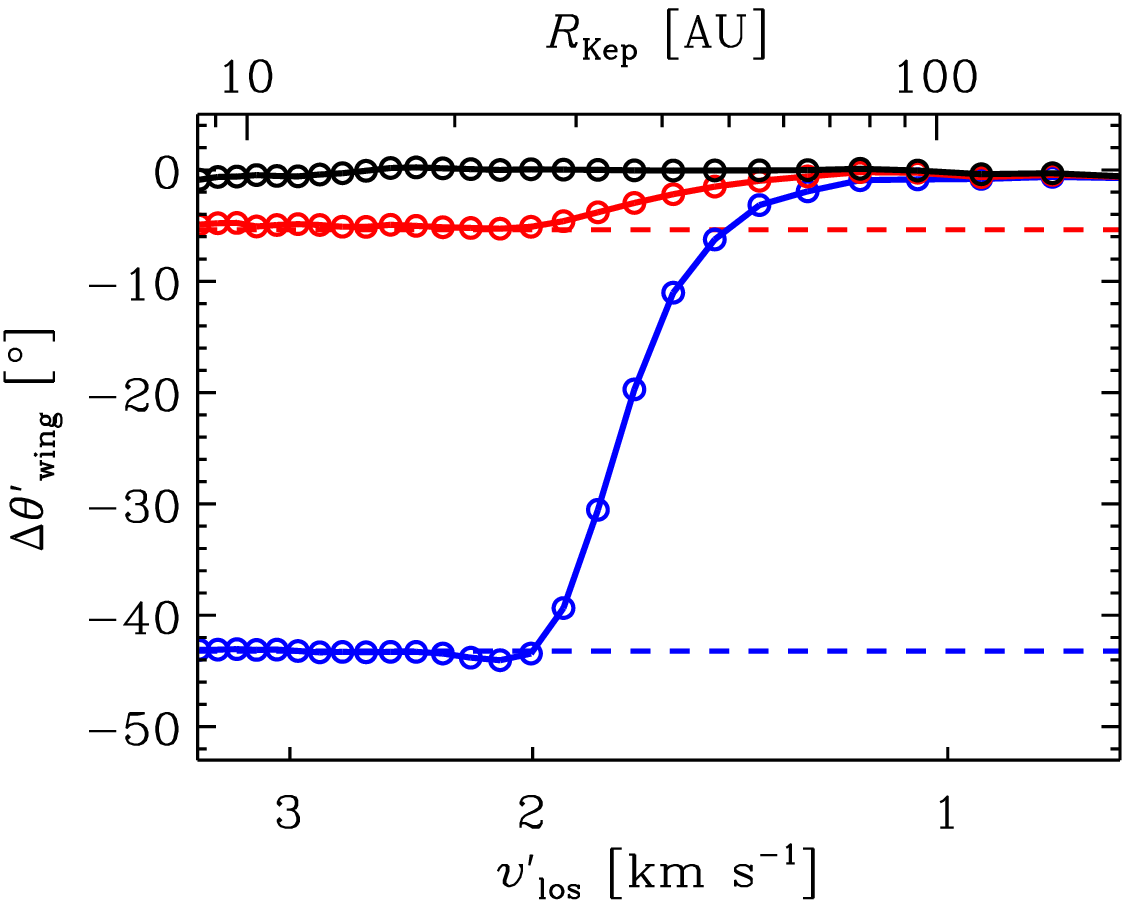}
\figcaption{Rotation angles $\Delta \theta'_{\rm sys}$ ({\it left}) and $\Delta 
\theta'_{\rm wing}$ ({\it right}) for $\chi = 0$ (black), 0.1 (red), and 1 
(blue) in 3D disk models with $i=20^\circ$ and $R_{\rm cav} = 50$ AU.  $\Delta 
\theta'_{\rm sys}$ is measured in the systemic velocity channel ($v'_{\rm los} 
= 0$) as a function of radius, and $\Delta \theta'_{\rm wing}$ is derived 
against channel velocity.  Analytic predictions from razor-thin disk models are 
drawn with horizontal dashed lines.  The lower abscissa of the left panel plots 
projected distance along the disk's major axis, while the upper abscissa of the 
right panel plots the radius for which $v_{\rm K} \sin i = v'_{\rm los}$.  
\label{fig:vsys}}
\end{figure}

\subsection{Twisted Isophotes at the Systemic Velocity} \label{subsec:isotwist}

The gas with no bulk motion relative to the systemic velocity of the disk has 
$v'_{\rm los} = 0$. From equation (\ref{eq:vlos}), this gas satisfies 
\begin{equation}
0=-v_{\rm K}\sin i\cos\theta_{\rm sys}+\chi v_{\rm K} \sin i\sin\theta_{\rm sys} \,,
\end{equation}
which implies that $\theta_{\rm sys}=\arctan(\chi^{-1})$ in the disk plane, and 
$\theta'_{\rm sys} = \arctan(\chi^{-1} \cos i)$ on the sky.  For $\chi = 0$ (no 
radial inflow), $\theta'_{\rm sys}=90^{\circ}$: the systemic velocity channel 
probes gas that lies along the minor axis of the disk as seen in projection 
(see Figure \ref{fig:models}, central panels).  When $\chi \neq 0$, the 
isovelocity contours at $v'_{\rm los} = 0$ will be rotated by an angle $\Delta 
\theta'_{\rm sys} = \theta'_{\rm sys} - 90^\circ$.  This rotation is evident in 
the top set of panels in Figure \ref{fig:thetasys}, which show the 
line-of-sight velocity fields of a razor-thin disk for various values of $\chi$ 
inside the disk cavity.  The rotation can also be seen in the isophotes 
observed at the systemic velocity of the corresponding 3D model; the bottom 
panels of Figure \ref{fig:thetasys} show that the isophotes in the channel maps 
``twist'' at $R = R_{\rm cav}$.

The isophote twist in the systemic velocity channel map can be
measured as a function of radius.  We perform this measurement by
first dividing the 3D disk model into concentric annuli, and further
splitting each annulus into its top ($y'>0$) and bottom ($y'<0$)
halves.  For each half-annulus, an intensity-weighted centroid (i.e.,
a ``center of brightness'') is computed.  The rotation angle $\Delta
\theta'_{\rm sys}$ is then evaluated from the line joining the
centroid positions of the top and bottom halves of a given annulus at
radius $R$.  The left panel of Figure \ref{fig:vsys} shows $\Delta
\theta'_{\rm sys} (y')$ for three values of $\chi$.  Note how $\max
|\Delta \theta'_{\rm sys}|$ is accurately predicted by the razor-thin
model, at least for the example inclination angle $i = 20^\circ$.
\highlight{In the mock observations of the 3D model, the transition
  between $\Delta\theta_{\rm sys}' = 0$ and $\max |\Delta\theta'_{\rm
    sys}|$, occurring near the cavity radius ($R_{\rm cav} = 50$\,AU),
  is smeared by the 0\farcs1 beam.  How well we can measure
  $\Delta\theta'_{\rm sys}$ in practice depends on the ratio of the
  beam size, $\theta_{\rm beam}$, to the peak signal-to-noise, $S/N$
  \citep[][]{reid88}. We estimate that a robust measurement of the
  twist requires $\theta_{\rm beam} \lesssim (S/N)\,\,|(R_{\rm
    cav}/d)\sin{i}\tan{\Delta\theta'_{\rm sys}}|$. For the 0\farcs1
  beam size adopted in Figure \ref{fig:vsys}, this requirement is met
  with $S/N > 10$ for $\chi = 0.1$, or with any detection of the line
  for $\chi = 1$.  Any measurement of an isophote twist from a
  synthesized image should be accompanied by a model imaged using the
  same $u,v$ sampling as the data, since the position angle of the
  resolution element can significantly alter the appearance of the
  image \citep{guilloteau98}.}

\begin{figure}[t!]
\epsscale{1.00}
\plotone{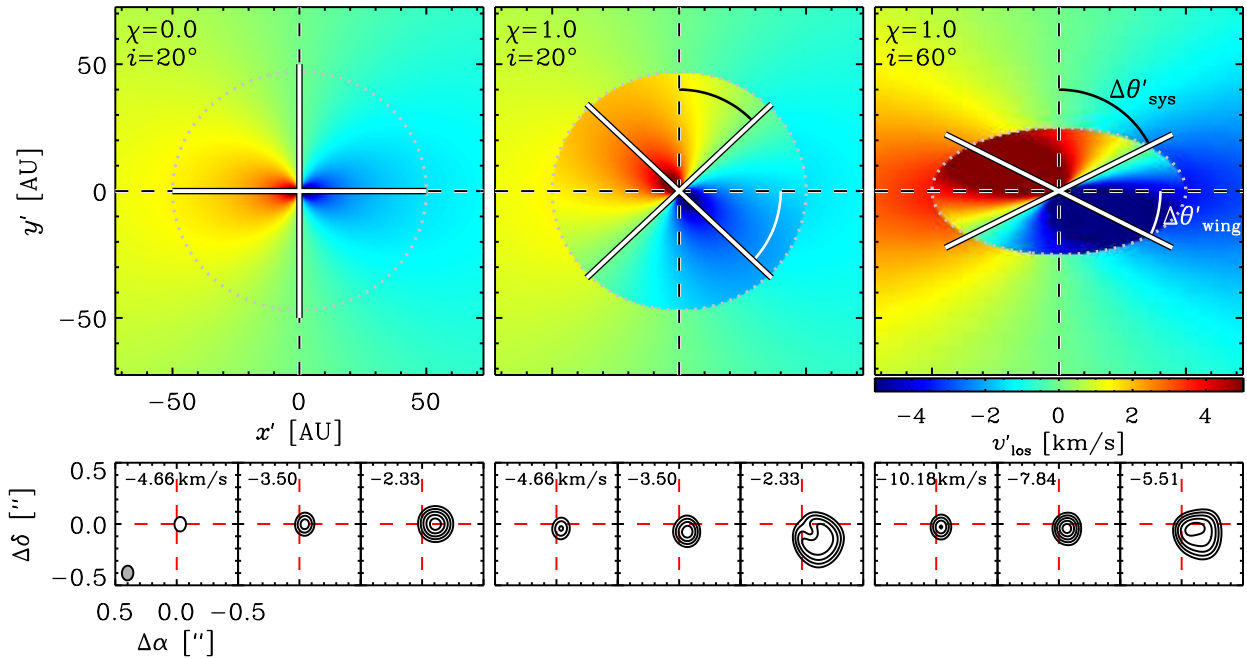}
\figcaption{Wing rotation. {\it Top}: Rotation angles $\Delta
\theta'_{\rm wing}$ (white arcs), overlaid on velocity fields of razor-thin 
disk models.  Other symbols are the same as in Figure \ref{fig:thetasys}.  Note 
how $\Delta \theta'_{\rm wing} \neq \Delta \theta'_{\rm sys}$.  {\it Bottom}: 
Simulated channel maps for various line-of-sight velocities from corresponding 
3D disk models.  Comparison of $\chi=0$ with $\chi=1$ shows how wing emission 
brightens and rotates clockwise from the projected major axis. 
\label{fig:thetawing}}
\end{figure}

\subsection{Maximum LOS Velocities: Wing Rotation} \label{subsec:wingrot}

\begin{figure}[t!]
\epsscale{0.50}
\plotone{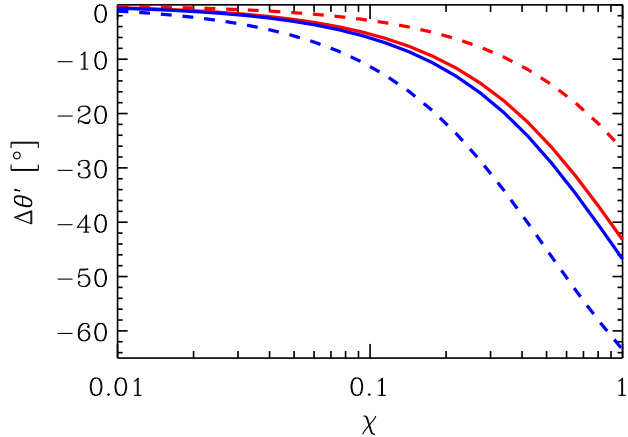}
\figcaption{Rotation angles $\Delta \theta'_{\rm sys}$ (blue) and $\Delta 
\theta'_{\rm wing}$ (red) for razor-thin disk models as a function of the 
radial inflow parameter $\chi$.  Solid and dashed curves are for disk 
inclinations $i = 20^\circ$ and 60$^\circ$, respectively.  At fixed $\chi$ and 
$i$, $|\Delta \theta'_{\rm sys}| > |\Delta \theta'_{\rm wing}|$.  Measuring 
$\Delta \theta'_{\rm wing}$ is easier at low inclinations, while measuring 
$\Delta \theta'_{\rm sys}$ is easier at high inclinations. 
\label{fig:comparetheta}}
\end{figure}

We now consider gas moving at the highest line-of-sight velocities, responsible 
for emission in the spectral line wings.  Maximizing equation (\ref{eq:vlos}) 
with respect to $\theta$ yields
\begin{equation}
0=v_{\rm K}\sin i\sin\theta_{\rm wing}+\chi v_{\rm K}\sin i\cos\theta_{\rm wing} \,,
\label{eq:wingrot}
\end{equation}
or $\theta_{\rm wing} =\arctan(-\chi)$ in the disk plane and $\theta'_{\rm 
wing} = \arctan(-\chi \cos i)$ on the sky.  Radial inflow causes the maximum 
isovelocity contour to rotate away from the disk major axis by $\Delta 
\theta'_{\rm wing} = \theta'_{\rm wing} (\chi) - \theta'_{\rm wing} (\chi = 0) 
= \theta'_{\rm wing} (\chi)$.  Figure \ref{fig:thetawing} shows the velocity 
fields for three razor-thin disk models, together with simulated channel maps 
for the corresponding 3D models.  Where there is inflow, emission from the 3D 
model is systematically rotated away from the disk major axis.  For extreme 
values of $\chi \sim 1$, line-of-sight velocities within the cavity are 
noticeably greater than for the case with no inflow (see top panels of Figure 
\ref{fig:thetawing}).  Correspondingly, emission from the line wings is 
brighter --- compare channels of the same velocity between the $\chi = 0$ and 
$\chi = 1$ models at fixed $i = 20^\circ$ (bottom panels of Figure 
\ref{fig:thetawing}) --- although it should be remembered that any decrease in 
the surface density inside the cavity (not modeled here) may reduce the 
intensity enhancement.  The last panel of Figure \ref{fig:thetawing} 
demonstrates that $\Delta\theta'_{\rm wing} \neq \Delta\theta'_{\rm sys}$.  
This difference is further quantified in Figure \ref{fig:comparetheta}, which 
shows that $|\Delta\theta'_{\rm wing} -\Delta\theta'_{\rm sys}|$ increases with 
increasing $i$ and $\chi$.  \highlight{Note how $|\Delta\theta'_{\rm wing}|$ is 
reduced for highly inclined disks, just opposite to the behavior of 
$|\Delta\theta'_{\rm sys}|$.  
Therefore, measuring $\Delta\theta'_{\rm wing}$ is easier 
at low inclinations, while measuring $\Delta\theta'_{\rm sys}$ is easier at high 
inclinations.  Clearly, disk orientation is a consideration when designing 
observations to detect either rotational signature.}

To measure $\Delta \theta'_{\rm wing}$ for the 3D disk model, we
employ a procedure similar to the one we devised for $\Delta
\theta'_{\rm sys}$.  For each spectral channel, we measure the
position of the intensity-weighted centroid (fitting two-dimensional
Gaussians produces consistent results).  Pairs of these positions are
constructed from channels blue- and red-shifted by the same velocity
relative to the systemic channel.  For a given velocity pair, the
angle $\Delta \theta'_{\rm wing}$ is calculated from the line segment
joining the two centroid positions.  The second panel in Figure
\ref{fig:vsys} shows $\Delta \theta'_{\rm wing}$ as a function of
channel velocity for a few values of $\chi$.  Rotation is evident for
all channel velocities equal to the Kepler velocities of material
within the cavity.  As was the case for $|\Delta \theta'_{\rm sys}|$,
the maximum value of $|\Delta \theta'_{\rm wing}|$ measured in the 3D
model is accurately predicted by the razor-thin model. \highlight{
  Measuring $|\Delta \theta'_{\rm wing}|$ in practice requires
  $\theta_{\rm beam}/(S/N) \lesssim |(R_{\rm cav}/d) \sin{i}
  \tan{\Delta\theta'_{\rm wing}}|$. For the model disks shown in Figure
  \ref{fig:thetawing} with $\chi = 1$, this beam size requirement is met when
  $S/N \ge 5$ and with relatively large beam sizes, $<$0\farcs5 or
  $<$0\farcs7 for $i = 20\degr$ or 60\degr, respectively.}

\highlight{In principle, the inclination $i$ and inflow parameter $\chi$ can 
be calculated if both $\theta'_{\rm sys}$ and $\theta'_{\rm wing}$ are 
measured.  In this case the product $\tan(\theta'_{\rm wing}) \tan(\theta'_{\rm 
sys}) = -\cos^2i$ and the ratio $\tan(\theta_{\rm wing})/\tan(\theta_{\rm sys}) 
= - \chi^2$.  The disk inclination so derived could then be compared to 
independent measurements.}

\highlight{These observational 
signatures of radial inflow --- twisted isophotes ($\Delta\theta'_{\rm sys}$) 
and wing rotation ($\Delta\theta'_{\rm wing}$) --- are intrinsically 
kinematic features, and therefore should be present regardless of which 
spectral line is used as a tracer of the projected disk velocity field.  That 
said, there is an obvious advantage to using lines with higher optical depths 
(e.g., $^{12}$CO, HCO$^+$), since they are more likely to be bright enough
inside the low-density cavities of transitional disks 
\citep[e.g.,][]{bruderer13} to enable robust measurements of radial inflow.}

\subsection{Disentangling Radial Inflow from Other Non-Keplerian Effects}
\label{subsec:disentangle}

The rotational signatures of inflow identified in \S\ref{subsec:isotwist} and 
\S\ref{subsec:wingrot} can be mimicked to varying degrees by other phenomena.  
We discuss four possible ``contaminants'' and how one might disambiguate 
between these different scenarios.

\subsubsection{Radial outflow} \label{out} 

Molecular gas can be entrained in winds flowing up and radially outward from 
the disk surface \citep[e.g.,][]{pontoppidan11,bast11,klaassen13}.  Radial 
outflow would cause the systemic velocity isophotes and the wing emission to 
rotate in the direction opposite to that for radial inflow.  Distinguishing 
inflow from outflow requires that we resolve the spin orientation of the disk; 
i.e., we need to decide whether $i < 90^{\circ}$ or $i > 90^{\circ}$.  The way 
to break this degeneracy is to use spatially resolved channel maps to decide 
which limb is approaching and which limb is receding, and then to combine this 
information with some asymmetry along the projected minor axis.  The asymmetry 
could be manifested in the orientation of a polar jet pointed toward the 
observer (i.e., whether the jet points along the positive $y > 0$ or negative 
$y < 0$ minor axis).  \highlight{Alternatively, one could observe, at any 
wavelength, a spectral line \citep{dutrey98,guilloteau98,rosenfeld13} or 
continuum \citep{augereau99,weinberger99,fukagawa06,thalmann10} brightness 
asymmetry arising from the disk's orientation.}  Of course it is also possible 
that one molecular line traces an outflowing disk wind, say because it probes 
gas at high altitude above the midplane, while another traces radial inflow, 
closer to the midplane.

\subsubsection{Infall from a molecular envelope}

Gas from the parent molecular cloud may continue to accrete onto the 
protoplanetary disk at early times \citep{stahler94,yorke99,brinch07}.  
Differential rotation and accelerating infall will mimic the spatio-kinematic 
signatures of inflow \citep[cf.][]{brinch08,tang12}.  However, the spatial 
scale of an infalling envelope should be comparable to or larger than the disk 
itself, and should not be confined to a central cavity.  Furthermore, for an 
isolated source, the duration of infall should be considerably shorter than the
lifetime of the disk \citep[e.g.,][]{pietu05}.  Evidence for local cloud 
material might be indicated by high extinction values, very red infrared 
spectral energy distributions, or contamination (self-absorption and/or 
emission on much larger angular scales) in single dish molecular line spectra.  
Most of the currently known transition disks do not exhibit any clear evidence 
for a remnant envelope structure.

\subsubsection{Vertical disk structure} \label{vert}

Line emission from optically thick gas typically originates in disk surface 
layers located at large vertical heights above the midplane
\citep[e.g.,][]{vanzadelhoff01,dartois03,pavlyuchenkov07,semenov08}.
Therefore, in channel maps, the blue- and red-shifted emission structures are 
not generally collinear with the stellar position; rather, they appear rotated 
away from the midplane in opposite directions from one another
\citep{rosenfeld13,degregorio-monsalvo13}.  These rotations of the blue- and 
red-shifted line wings would effectively cancel out in the procedure we 
developed in \S \ref{subsec:wingrot} to measure $\Delta \theta'_{\rm wing}$.
Moreover, vertical structure does not twist the isophotes at the systemic 
velocity (i.e., $\Delta \theta'_{\rm sys} = 0$).  In sum, vertical structure 
should pose little danger of confusion in efforts to characterize radial 
inflow.

\begin{figure}[t!]
\epsscale{1.0}
\plottwo{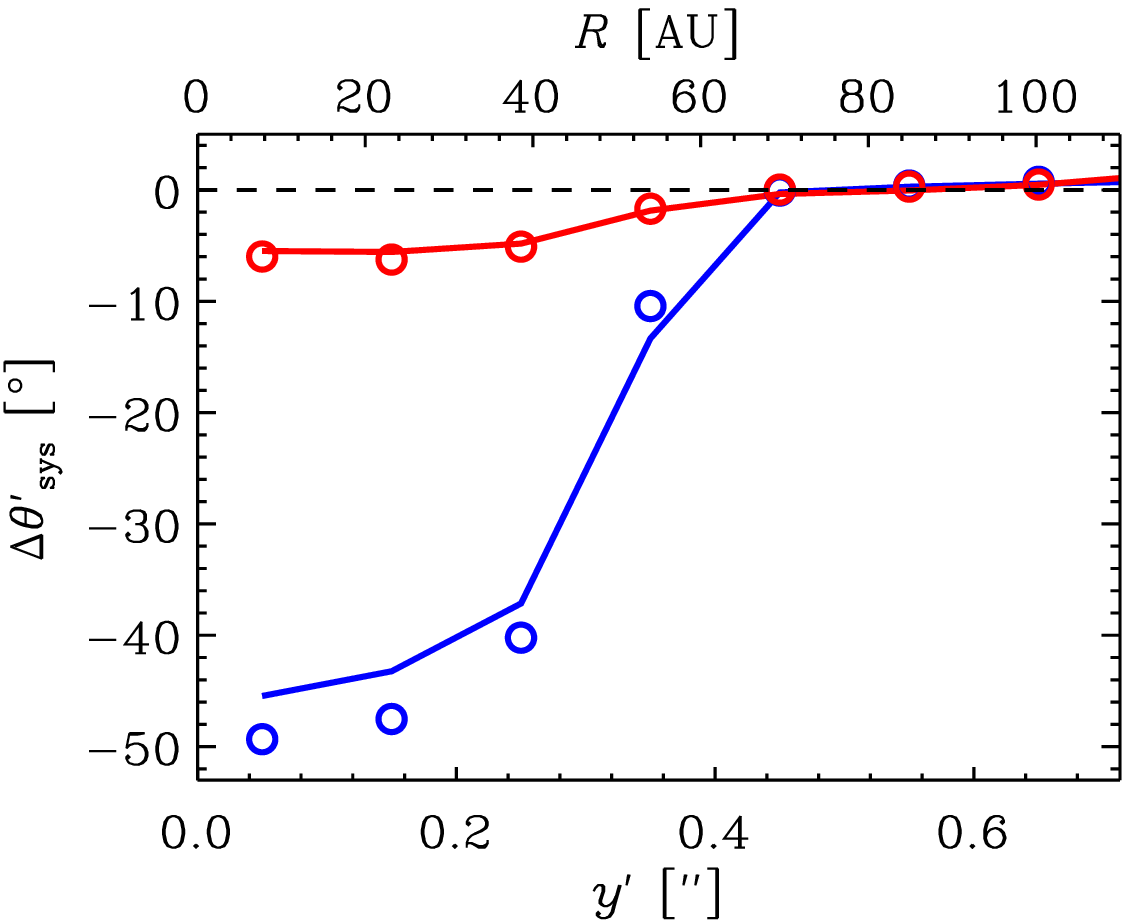}{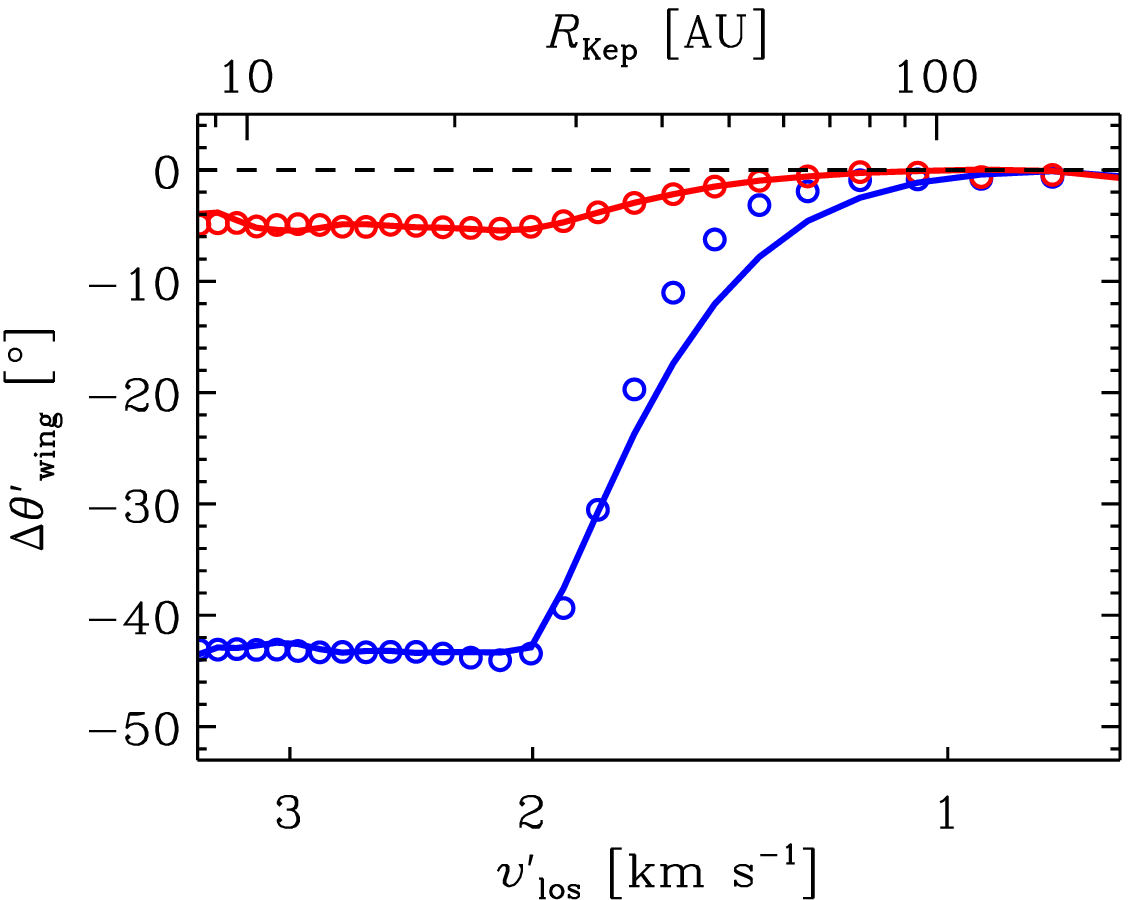}
\figcaption{Degeneracy between a warp and radial inflow.  The same rotation 
angles for inflow models as shown in Figure \ref{fig:vsys} are plotted here as 
open circles, and overlaid with solid curves generated from model warped 
disks.  Warp parameters were chosen to reproduce $\Delta \theta_{\rm wing}'$ 
for the inflow models; evidently these same warp models automatically mimic 
$\Delta \theta_{\rm sys}'$ as well.  For $\chi=1$, the warp parameters are $R_w 
= R_{\rm cav} = 50$\,AU, $i_{\rm rel}=14^\circ$, and $\Omega_{\rm rel} = 
250\degr$.  For $\chi = 0.1$, $R_w = 50$\,AU, $i_{\rm rel} = 2\degr$, and 
$\Omega_{\rm rel}=267\degr$.  \label{fig:vsys_warp}}
\end{figure}

\subsubsection{A warp}\label{warp}

A warp --- introduced, for example, by a perturbing body whose orbit
is inclined to the disk plane \citep{larwood97,marzari09} ---
\highlight{also changes the morphology of line emission in channel
  maps. We have found that, unfortunately, the warp-induced changes in the channel maps are practically identical to
  those induced by a radial inflow.}\footnote{\highlight{We also
    tested whether disk eccentricity was degenerate with the signature of
    radial inflow, and found that it was not (data not shown).}}  
At its most basic, a warp comprises an inner disk that is misaligned with
an outer disk, and is described by three parameters: the radius $R_w$
where the inner disk transitions to the outer disk, the inclination
$i_{\rm inner}$ of the inner disk relative to the sky plane, and the
orientation of the inner disk, i.e., the longitude $\Omega_{\rm
  inner}$ of ascending ($\equiv$ toward the observer) node, measured
in the sky plane and referred to the $x=x'$ axis (as a reminder, the
outer disk has inclination $i_{\rm outer} \equiv i$ relative to the
sky plane and orientation $\Omega_{\rm outer} \equiv 0^\circ$).  An
equivalent description uses the outer disk as the reference plane;
under this alternate convention, the inner disk inclination relative
to the outer disk is $i_{\rm rel}$, and its longitude of ascending
node is $\Omega_{\rm rel}$, measured in the outer disk plane and
referred to the $x=x'$ axis.  The angles in the two reference systems
are related by
\begin{align}
i_{\rm rel}=&\arccos\left( \cos i \cos i_{\rm inner}+
    \cos \Omega_{\rm inner} \sin i \sin i_{\rm inner} \right) \\
\Omega_{\rm rel} =& \frac{\pi}{2} + 
    \arctan\left(\frac{\cos i_{\rm inner} \sin i - 
    \cos i \cos \Omega_{\rm inner} \sin i_{\rm inner} }{\sin i_{\rm inner} 
    \sin \Omega_{\rm inner} } \right) \,.
\end{align}

Figure \ref{fig:vsys_warp} shows $\Delta\theta'_{\rm sys}$ and 
$\Delta\theta'_{\rm wing}$ for two 3D disk models with warps (solid curves) 
overlaid on our previous results for disk models with radial inflow (open 
circles).  The similarity between these rotation angle patterns demonstrates 
that a warp with the right parameters (see caption) can masquerade as inflow 
\highlight{in channel maps}.  Furthermore, increasing 
$i_{\rm rel}$ can brighten wing emission in the same way that increasing $\chi$ 
does \citep[data not shown; see \S \ref{subsec:wingrot};][]{rosenfeld12}. 

Distinguishing inflow from a warp requires additional data and modeling.  A 
warped disk located in a region that is known to be optically thin at continuum 
wavelengths would need to have its opacity be reduced by large factors by, 
e.g., grain growth or dust filtration at the cavity rim.  By contrast, in the 
fast inflow model, the opacity need not change much, if at all, since the 
surface density reduction that accompanies fast inflow would account for most 
if not all of the cavity's transparency.  Spatially resolved, multi-wavelength 
imaging of the disk cavity can serve to constrain changes in the dust size
distribution.  And one can utilize emission lines from rarer, optically thin 
gas species (e.g., \citealt{bruderer13}) to measure total gas surface 
densities.  In \S\ref{sec:prospects} we wrestle again with the warp/inflow 
degeneracy using real observations.\footnote{A warp and inflow are not mutually 
exclusive possibilities --- the two phenomena might be driven simultaneously by 
companions inside the disk cavity.}

\section{A CASE STUDY: HD 142527} \label{sec:prospects}

We consider the remarkable transition disk hosted by the young F star HD 
142527.  At a distance of $d=145 \pm 15$ pc \citep{verhoeff11}, the disk has a 
large dust-depleted cavity of radius $R_{\rm cav} \sim 130$ AU that has been 
imaged in near-infrared scattered light 
\citep{fukagawa06,casassus12,rameau12,canovas13} and mid-infrared thermal 
emission \citep{fujiwara06,verhoeff11}.  The cavity is not completely devoid of 
dust; \citet{verhoeff11} required an optically thick disk (and even an 
optically thin halo) interior to $\sim$30 AU to explain the observed 
near-infrared and mid-infrared emission.  ALMA Band 7 (345\,GHz) observations 
presented by \citet{casassus13} resolved the highly asymmetric dust ring at 
$R_{\rm cav}$ that was originally noted by \citet{ohashi08}.  The ALMA data 
clearly demonstrate that molecular gas resides inside the cavity, emitting in
the $^{12}$CO $J$=3$-$2 and HCO$^+$ $J$=4$-$3 lines.  We have already noted the 
nearly free-fall radial velocities implied by the HCO$^+$ filament 
(\S\ref{intro}).

\begin{figure}[t!]
\epsscale{1.00}
\plotone{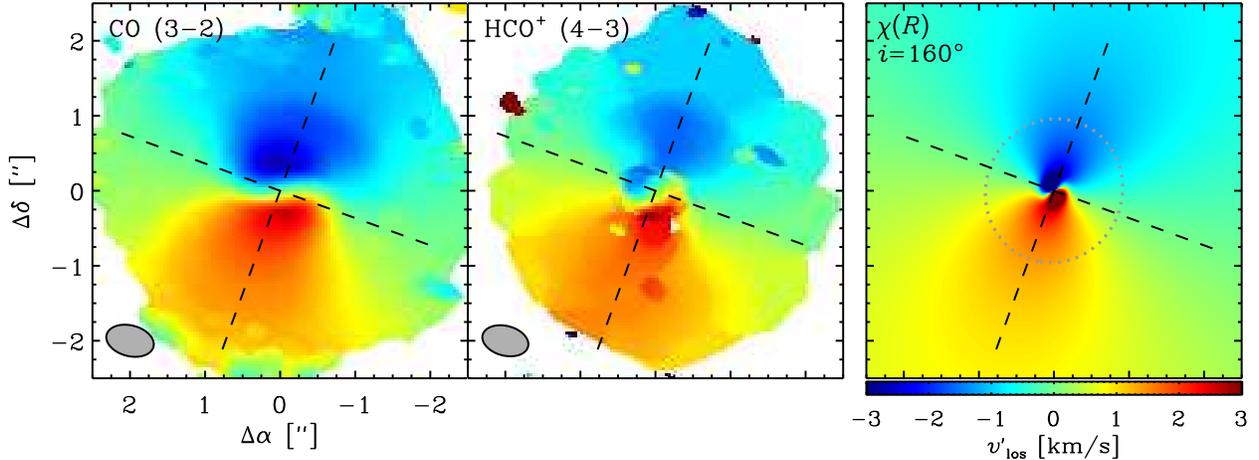}
\figcaption{Evidence for fast radial inflow in the HD 142527 disk.  {\it Left:} 
First moment or intensity-weighted velocity map of the $^{12}$CO $J$=3$-$2 
emission line constructed from archival ALMA data.  {\it Center:} First moment 
map of the HCO$^+$ $J$=4$-$3 emission line.  The HCO$^+$ data has lower 
signal-to-noise than the CO data.  {\it Right:} Velocity field for a razor-thin 
disk with the inflow parameter $\chi$ increasing with decreasing $R$ and 
peaking at unity (see text for details).  All panels are colored using the same 
velocity scale, and the orientation of the outer disk is indicated by dashed 
black lines (projected major axis at PA = $-20\degr$).  The counter-clockwise 
twist, a signature of radial inflow for $i > 90\degr$, is present in all 
panels; in the first moment maps, the twist appears on larger scales where the 
emission line intensity is strongest.  \label{fig:hd142527_moms}}
\end{figure}

We apply the tools developed in \S\ref{sec:twist} to the archival ALMA data on 
CO and HCO$^+$ line emission. We aim here only for a first look at the 
kinematics, and so forego an extensive modeling effort, fixing in advance the 
disk mass ($0.1\,M_\odot$), scale radius ($R_s = 50$ AU), disk temperature ($T 
= 70 (R/50 \,{\rm AU})^{-0.7}$\,K), and outer truncation radius (300 AU).  We also 
adopt literature values for the stellar mass \citep[$M = 
2.7\,M_\odot$;][]{casassus13} and viewing geometry for the outer disk 
\citep[$i = 160\degr$, with the projected major axis oriented at position 
angle PA = -20$\degr$;\footnote{Alternatively, if our $x'$-axis points west and 
our $y'$-axis points north, then $\Omega_{\rm outer} = 
70\degr$.}][]{fujiwara06,verhoeff11}. The inclination and PA of the outer disk
are based on CO channel maps together with thermal mid-infrared images which 
show that the eastern side of the disk appears brighter than the western side;
on the eastern side we are seeing the portion of the cavity rim located farther 
from the observer and directly illuminated by starlight.  The line-of-sight 
velocities reported below are all relative to systemic; the systemic velocity 
is +3.64\,km\,s$^{-1}$ relative to the local standard of rest, as we deduced 
from visual inspection of the CO channel maps.

Figure \ref{fig:hd142527_moms} shows the first moment maps for the CO
and HCO$^+$ emission lines.  The first moment or intensity-weighted
line-of-sight velocity appears twisted near the disk center ($R'
\lesssim 0\farcs5$).  With respect to the orientation defined by the
outer disk, the twist in the inner disk is counter-clockwise, which
given $i=160\degr$ implies radial inflow.  Furthermore, the twist is of order
$\sim$1 radian in magnitude, which implies a radial speed comparable to 
Keplerian ($\chi \sim 1$).  Compare the first moment maps (left 
and center panels) with the velocity field of a model razor-thin disk with a 
fast inflow (right panel; cf.~Figure \ref{fig:thetawing}), and recognize that 
the comparison is only meant to be suggestive since the velocity field 
is not the same as the intensity-weighted velocity.

\begin{figure}[t!]
\epsscale{0.50}
\plotone{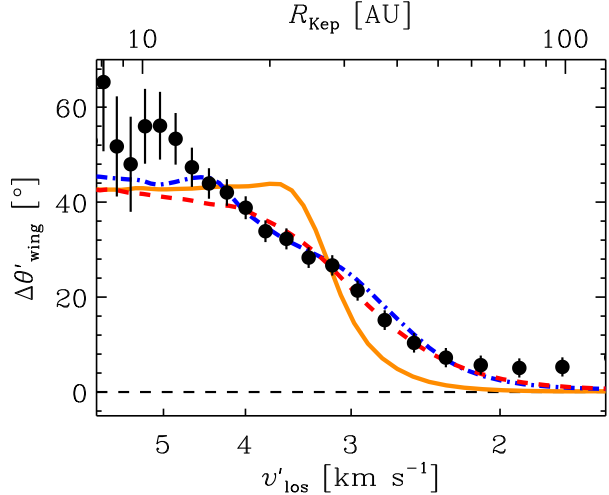}
\caption{Possible fast inflow in the HD 142527 disk.  The measured rotation,
$\Delta\theta'_{\rm wing}$, in the $^{12}$CO $J$=3$-$2 line wings can be 
approximately reproduced by either a disk with graded radial
inflow (spatially variable $\chi$; red dashed curve), or a graded warp 
(spatially variable $i_{\rm rel}$ and $\Omega_{\rm rel}$; blue dot-dashed 
curve).  Details of the 
models are in the main text.  An inflow model having constant $\chi = 1$ for $R 
< 35$ AU (orange) does less well than the graded inflow model.  The spin 
orientation of the HD 142527 disk ($i = 160\degr$) is opposite to those of our 
example disks (having $i < 90\degr$) in \S\ref{sec:twist}, and so 
$\Delta\theta'_{\rm wing}$ is positive rather than negative.  
\label{fig:hd142527_rot}}
\end{figure}

Figure \ref{fig:hd142527_rot} compares the measured $\Delta\theta'_{\rm wing} 
(v'_{\rm los})$ in the CO line with predictions from several models.  A 
constant $\chi$ model does not fit the data as well as a variable $\chi (R < 
140 \, {\rm AU} = [1+(R/33$\,AU$)^6]^{-1}$ model that introduces the twist
incrementally.  The variable $\chi$ model does as well as a warped disk model 
(cf.~\S\ref{warp}) in which the warp is likewise introduced gradually.  To mock 
up a warp induced by a planet as described by linear secular theory 
\citep[e.g.,][]{dawson11}, we ramp up the relative inclination linearly from 0 
to its final value $i_{\rm rel} = 15\degr$, and ramp down the nodal angle 
linearly from $\Omega_{\rm rel} + 90\degr$ to $\Omega_{\rm rel} = 69\degr$, as 
the disk radius decreases from $2R_w$ to $R_w = 25$ AU.

At the highest velocities/smallest radii probed in Figure 
\ref{fig:hd142527_rot}, all our models systematically underpredict $\Delta 
\theta'_{\rm wing}$.
For the inflow model, the fit at small radii could be improved by having the 
azimuthal velocity decrease by a few tens of percent in tandem with the 
increase in radial inflow velocity.  The large synthesized beam size 
(0\farcs64$\times$0\farcs41, PA = 75.5$^\circ$) precludes us from constructing 
a plot of $\Delta \theta'_{\rm sys} (y')$ showing what would presumably be 
large twists at small radii (cf.~Figure \ref{fig:vsys}, which was made assuming 
a beam size of 0\farcs1).

At the smallest velocities/largest radii, there appears in Figure
\ref{fig:hd142527_rot} to be a constant offset in $\Delta \theta'_{\rm
  wing}$ of about 6 degrees. The offset may be due to a systematic
error in the position angle we assumed for the outer disk.  Another
source of error at these low velocities may be in our systemic
velocity (see \citealt{hughes11} for a method to quantify this
uncertainty).

To summarize, our first-cut analysis of the \citet{casassus13} ALMA 
observations of the HD 142527 disk demonstrates that the predicted 
spatio-kinematic signatures of fast radial inflow are eminently detectable, and 
merits a dedicated pursuit of inflow in a general sample of transition disks.  
Additional observations and modeling are posited to distinguish between a fast 
inflow or warp origin for the observed kinematics.  This particular disk is 
especially challenging to interpret because of its complex and non-axisymmetric 
structures \citep[][]{fukagawa06,fujiwara06,verhoeff11,casassus12,honda12,
rameau12,casassus13,canovas13}.  The disk interior to $R\sim 30$ AU that was 
inferred by \citet{verhoeff11} could, in principle, support a warp.  
Observations of an optically thin line that resolve the central cavity would 
help clarify the situation, to confirm the lower gas surface densities that 
should accompany fast inflow.  \highlight{Indeed, a recent study of the 
$^{13}$CO and C$^{18}$O $J$=3$-$2 lines in this disk by \citet{fukagawa13} 
support the inflow scenario: they detect little (if any) 
emission from these optically thin isotopologues inside the disk cavity.  No 
evidence for twisted isophotes or wing rotation is noted by \citet{fukagawa13} in these 
lines, but that is not surprising, as the emission is too faint --- the current S/N is too 
low --- inside the cavity region to detect these features.}

\section{CONCLUSIONS}\label{sec:conclude}

Radial inflow of gas at velocities approaching free fall can account for both 
the depleted densities inside transition disk cavities and the relatively 
normal stellar accretion rates they maintain.  We have demonstrated that high 
spatial and spectral resolution observations of molecular emission lines with
(sub)millimeter wavelength interferometers --- particularly ALMA --- can detect 
inflow by virtue of its two key signatures: twisted isophotes at the systemic 
velocity, and rotated emission patterns in channel maps made in the line wings.
We developed quantitative diagnostics of channel maps to characterize radial 
inflow, and discussed various real-world complications.  Of these, the most 
serious is a warp.  

We tested our inflow diagnostics on archival ALMA observations of the HD 142527
transition disk, uncovering a booming signal consistent with free-fall inflow 
velocities at distances of $\sim$25--50 AU from the host star.  Nevertheless a 
warped disk at these radii can also reproduce the large, order-unity rotations 
seen in the CO and HCO$^+$ line wings.  Whether the HD 142527 disk cavity 
contains a fast radial inflow or a warped disk (or both), the only mechanism on 
the table now that can explain either phenomenon involves strong gravitational
torques, exerted by one or more giant planets/brown dwarfs/low-mass stars 
\citep[as yet undetected; see][]{biller12,casassus13}.  Fast flows should also 
produce strong shocks in disks.  We have not investigated the signatures of 
such shocks; the observables depend on details such as the density
and the cooling mechanisms of shocked material.
In the central cavity of the circumbinary disk around GG Tau A,
\cite{beck12} have detected hot H$_2$
that may trace shocks internal to the accretion flow.

\acknowledgments We thank Ruobing Dong, Eve Lee, and Scott Tremaine
for discussions, and the organizers of the 2013 Gordon Conference on
Origins of Solar Systems, Ed Young and Fred Ciesla, for providing a
great venue for sparking collaboration on this article.  We also thank
the referee Takayuki Muto for many insightful
comments.  This paper uses the following ALMA data:
ADS/JAO.ALMA\#2011.0.00465.S.  ALMA is a partnership of ESO
(representing its member states), NSF (USA) and NINS (Japan), together
with NRC (Canada) and NSC and ASIAA (Taiwan), in cooperation with the
Republic of Chile.  The Joint ALMA Observatory is operated by ESO,
AUI/NRAO and NAOJ.  The National Radio Astronomy Observatory is a
facility of the National Science Foundation operated under cooperative
agreement by Associated Universities, Inc.

\bibliography{shortbib_1206}

\end{document}